\ifx\pdfsuppresswarningpagegroup\undefined\else\pdfsuppresswarningpagegroup1\fi
\documentclass[a4paper,11pt]{article}
\usepackage{pos}
\usepackage{amsfonts} 
\usepackage{amssymb} 
\usepackage{amsmath} 
\usepackage{graphicx} 
\graphicspath{{..}}
\usepackage{subfigure} 
\usepackage{array} 
\usepackage{dcolumn} 
\usepackage{bm} 
\let\bold=\bm 
\usepackage{latexsym} 
\usepackage{longtable} 
\usepackage{hyperref} 
\usepackage{bm}
\usepackage{slashed}
\usepackage{bbold}
\usepackage{color}
\usepackage{multirow}
\usepackage{rotating}
\usepackage{comment}
\usepackage [english]{babel}
\usepackage [autostyle, english = american]{csquotes}
\MakeOuterQuote{"}
\usepackage{graphicx}
\usepackage{subfigure}
\usepackage{xcolor}
\usepackage{appendix}[toc,page]
\usepackage{setspace}
\usepackage{amsmath, amssymb}
\usepackage{slashed}
\newcommand{\be}{\begin{eqnarray}}
\newcommand{\ee}{\end{eqnarray}}

\newcommand{\gf}{\gamma^5}
\newcommand{\ra}{\rangle}
\newcommand{\la}{\langle}

\providecommand{\abs}[1]{\lvert#1\rvert}
\providecommand{\matrixe}[3]{\langle#1\lvert#2\rvert#3\rangle}

\def\ol{\overline}

\makeatletter
\newcommand{\overleftrightsmallarrow}{\mathpalette{\overarrowsmall@\leftrightarrowfill@}}
\newcommand{\overrightsmallarrow}{\mathpalette{\overarrowsmall@\rightarrowfill@}}
\newcommand{\overleftsmallarrow}{\mathpalette{\overarrowsmall@\leftarrowfill@}}
\newcommand{\overarrowsmall@}[3]{%
  \vbox{%
    \ialign{%
      ##\crcr
      #1{\smaller@style{#2}}\crcr
      \noalign{\nointerlineskip}%
      $\m@th\hfil#2#3\hfil$\crcr
    }%
  }%
}
\def\smaller@style#1{%
  \ifx#1\displaystyle\scriptstyle\else
    \ifx#1\textstyle\scriptstyle\else
      \scriptscriptstyle
    \fi
  \fi
}
\makeatother
\newcommand{\olra}[1]{\overleftrightsmallarrow{#1}}

\title{Nucleon isovector momentum fraction, helicity and transversity moment using Lattice QCD}
\ShortTitle{Nucleon isovector moments using Lattice QCD}

\author*[a,c]{Santanu Mondal}
\author[a]{Tanmoy Bhattacharya}
\author[a]{Rajan Gupta}
\author[b]{B\'alint~Jo\'o}
\author[c,d]{Huey-Wen Lin}
\author[a]{Sungwoo Park}
\author[e]{Frank Winter}
\author[f]{Boram Yoon}

\affiliation[a]{Los Alamos National Laboratory, Theoretical Division T-2, Los Alamos, NM 87545}

\affiliation[b]{Oak Ridge Leadership Computing Facility, Oak Ridge National Laboratory, Oak Ridge, TN 37831, USA}

\affiliation[c]{Department of Physics and Astronomy, Michigan State University, MI, 48824, USA}
\affiliation[d]{Department of Computational Mathematics, Science and Engineering, Michigan State University, MI, 48824, U.S.A}

\affiliation[e]{Jefferson Lab, 12000 Jefferson Avenue, Newport News, Virginia 23606, USA}
\affiliation[f]{Los Alamos National Laboratory, Computer Computational and Statistical Sciences, CCS-7, Los Alamos, NM 87545}

\emailAdd{mondals2@msu.edu}

\abstract{We present our recent high precision calculations \cite{Mondal:2020cmt, Mondal:2020ela} of the first moment of nucleon isovector polarized, unpolarized and transversity distributions, i.e., momentum fraction, helicity and transversity moment, respectively. 
We use the standard
method for the calculation of these moments (via matrix elements of twist two operators),
and carry out a detailed analysis of the sources of systematic uncertainty, in particular of excited state contributions. Our calculations have been performed using two different lattice setups (Clover-on-HISQ and
Clover-on-Clover), each with several ensembles. They give consistent results that are in agreement
with global fit analyses.}

\FullConference{%
 The 38th International Symposium on Lattice Field Theory, LATTICE2021
  26th-30th July, 2021
  Zoom/Gather@Massachusetts Institute of Technology}

\begin{document}
\maketitle

\section{Introduction}
In the realm of QCD, among the key quantities to quantitatively characterize the rich and complex structure of hadrons are a number of universal,
non-perturbative distribution functions.
These are, parton distribution functions (PDFs),
transverse momentum dependent PDFs (TMDs), generalized parton distributions (GPDs)
and distribution amplitudes (DAs). For many years, there have been steady efforts to obtain these distributions both from experiments and theory. On the theoretical side, information on the moments of the distribution functions can be obtained via first principle Lattice QCD calculations. Subsequent to the proposal by Ji in 2013 \cite{Ji:2013dva}, there have also been significant progress towards accessing the distributions themselves on the lattice \cite{Cichy:2021lih}.

The distributions are not measured directly in experiments, and phenomenological analyses including different theoretical inputs are needed to extract them from experimental data.
In cases where both lattice results and phenomenological analyses of experimental data (global fits) exist, one can compare them to validate the control over systematics in the lattice calculations, and on the other hand provide a check on the phenomenological process used to extract these observables from experimental data~\cite{Lin:2017snn, Constantinou:2020hdm}. In other cases, lattice results are predictions.

Even for the best studied quantity on the lattice, the isovector momentum fraction $\la x \ra_{u-d}$, the data had large statistical and systematic uncertainties prior to 2018 \cite{Lin:2017snn}. Here we present high-precision lattice calculations for the isovector momentum fraction, helicity moment $\la x \ra_{\Delta u- \Delta d}$ and transversity moment $\la x \ra_{\delta u-\delta d}$ using two different lattice setups and controlling the sources of systematic uncertainties.  
Our study shows, that the lattice data for these three moments are now of quality comparable to that for nucleon charges (zeroth moments).

\section{Lattice set up}

\begin{table}
\begin{tabular}{ |c|c|c|c|c|c|c| }
\hline
Ensemble&$a$&$M_\pi$& $L^3\times T$&$M_\pi L$&$\tau/a$  &$N_{conf}$\\
ID &(fm)&(MeV)&&&&\\
\hline
\hline
$a15m310$&$0.1510(20)$&$320.6(4.3)$&$16^3\times48$&$3.93$&$\{5,6,7,8,9\}$    &$1917$       \\
\hline
$a12m310$&$0.1207(11)$&$310.2(2.8)$&$24^3\times64$&$4.55$&$\{8,10,12,14\}$   &$1013$        \\
$a12m220$&$0.1184(09)$&$227.9(1.9)$&$32^3\times64$&$4.38$&$\{8,10,12,14\}$   &$1156$        \\
$a12m220L$&$0.1189(09)$&$227.6(1.7)$&$40^3\times64$&$5.49$&$\{8,10,12,14\}$  &$1000$       \\

\hline
$a09m310$&$0.0888(08)$&$313.0(2.8)$&$32^3\times96$&$4.51$&$\{10,12,14,16\}$  &$2263$       \\
$a09m220$&$0.0872(07)$&$225.9(1.8)$&$48^3\times96$&$4.79$&$\{10,12,14,16\}$  &$960$        \\
$a09m130$&$0.0871(06)$&$138.1(1.0)$&$64^3\times96$&$3.90$&$\{10,12,14,16\}$  &$1041$        \\
\hline
$a06m310W$&$0.0582(04)$&$319.6(2.2)$&$48^3\times144$&$4.52$&$\{18,20,22,24\}$ &$500$        \\
$a06m135$&$0.0570(01)$&$135.6(1.4)$&$96^3\times192$&$3.7$&$\{16,18,20,22\}$  &$751$         \\
\hline
\hline
\end{tabular}
\caption{Lattice parameters of the $2+1+1$-flavor HISQ ensembles generated by the MILC collaboration~\cite{Bazavov:2012xda} and analyzed in this study. We give the lattice spacing $a$, pion mass $M_\pi$, lattice size $L^3 \times T$, the values of
source-sink separation $\tau$ simulated, and the number of configurations analyzed.}
\label{Table:HISQ-ensembles}
\end{table}

\begin{table}[tbhp]  

\centering
\renewcommand{\arraystretch}{1.2}
\begin{tabular}{|c|c|c|c|c|c|c| }
\hline
Ensemble&$a$&$M_\pi$& $L^3\times T$&$M_\pi L$&$\tau/a$
&$N_{conf}$\\
ID &  (fm)&(MeV)&&&&\\
\hline
\hline
$a127m285$&$0.127(2)$&$285(3)$&$32^3\times96$&$5.85$&$\{8,10,12,14\}$     &$2001$       \\
\hline
$a094m270$&$0.094(1)$&$270(3)$&$32^3\times64$&$4.11$&$\{10,12,14,16,18\}$ &$1464$        \\
$a094m270L$&$0.094(1)$&$269(3)$&$48^3\times128$&$6.16$&$\{10,12,14,16\}$ &$4501$        \\
\hline
$a091m170$&$0.091(1)$&$169(2)$&$48^3\times96$&$3.74$&$\{8,10,12,14,16\}$   &$4015$       \\
$a091m170L$&$0.091(1)$&$169(2)$&$64^3\times128$&$5.08$&$\{8,10,12,14,16\}$&$1533$       \\
\hline
\hline
$a073m270$&$0.0728(8)$&$272(3)$&$48^3\times128$&$4.8$&$\{11,13,15,17,19\}$&$4477$       \\
$a071m170$&$0.0707(8)$&$167(2)$&$72^3\times192$&$4.26$&$\{15,17,19,21\}$
&$2100$        \\
$a071m130$&$0.0707(8)$&$127(1)$&$96^3\times192$&$4.36$&$\{13,15,17,19,21\}$
&$440$        \\
\hline
$a056m280$&$0.056(1)$&$280(5)$&$64^3\times192$&$5.09$&$\{18,21,24,27,30\}$
&$1723$        \\
\hline
\end{tabular}
\caption{Lattice parameters of the $2+1$-flavor clover ensembles generated by the JLab/W\&M/LANL/MIT collaboration and analyzed in this study. We give the lattice spacing $a$, pion mass $M_\pi$, lattice size $L^3 \times T$, the values of source-sink separation $\tau$ simulated, and the number of configurations analyzed.}
\label{Table:Clover-ensembles}
\end{table}

We present our calculations of the three moments using two different lattice setups: (i) the Clover-on-HISQ calculation (PNDME 20) published in \cite{Mondal:2020cmt} was performed using nine HISQ ensembles generated by the MILC collaboration \cite{Bazavov:2012xda}, whose parameters are summarized in Table \ref{Table:HISQ-ensembles}. They cover a range of lattice spacings $(0.057 \leq a \leq 0.15)$ fm, pion masses $(135 \leq M_\pi \leq 310 )$ MeV and lattice sizes
$(3.7 \leq M_\pi L \leq 5.5)$. For more details of the lattice methodology, the strategies for the calculations and the analyses see \cite{Mondal:2020cmt} and references therein.

(ii) The Clover-on-Clover calculation, published in \cite{Mondal:2020ela} (NME 20), 
used seven Clover ensembles generated by the JLab/W\&M/LANL/MIT collaboration \cite{ensembles_generation:C-on-C}. Here (NME 21) we include two new ensembles, one at physical $M_\pi$ and the other at smaller lattice spacing $a = 0.056$ fm which gives us better control over chiral and continuum extrapolations, respectively. Note that the data for the two new ensembles are preliminary and  $a071m130$ is statistics limited. 
The parameters of these Clover ensembles are summarized in Table \ref{Table:Clover-ensembles}. They also cover a range of lattice spacings $(0.056 \leq a \leq 0.127)$ fm, pion masses $(127 \leq M_\pi \leq 285 )$ MeV and lattice sizes
$(3.74 \leq M_\pi L \leq 5.85)$.

We construct the correlation functions needed to calculate the matrix elements using Wilson-clover fermions for both lattice setups. The Clover-on-HISQ formulation is non-unitary and can suffer
from the problem of exceptional configurations at
small, but a priori unknown, quark masses.
However, we have not found evidence for such exceptional configurations on any of the nine ensembles analyzed in
this work.

\section{Lattice correlators and moments}

The light quark operators ($q \in \{u,d\}$) used to calculate the moments are:
\be
\la x \ra_{u-d}&:& {\cal O}^{44}_{V^3} =  \ol{q} (\gamma^{4}\olra{D}^{4}  -\frac{1}{3}
{\bm \gamma} \cdot \olra{\bf D}) \tau^3 q
\label{eq:finaloperatorV} \\
%
%
\la x \ra_{\Delta u-\Delta d}&:& {\cal O}^{34}_{A^3}=\ol{q} \gamma^{\{3}\olra{D}^{4\}} \gf \tau^3 q
\label{eq:finaloperatorA} \\
\la x \ra_{\delta u-\delta d}&:& {\cal O}^{124}_{T^3}=\ol{q} \sigma^{[1\{2]}\olra{D}^{4\}} \tau^3 q \,.
\label{eq:finaloperatorT}
\ee
From their matrix elements within the ground state of the nucleon, the moments are given by:
\be
\la 0 | {\cal O}^{44}_{V^3}| 0 \ra &=&  -  M_N\, \la x \ra_{u-d} \,,
\label{eq:me2momentV} \\
\la 0 | {\cal O}^{34}_{A^3}| 0 \ra &=&  - \frac{i  M_N}{2} \, \la x \ra_{\Delta u-\Delta d} \,,
\label{eq:me2momentA} \\
\la 0 | {\cal O}^{124}_{T^3}| 0 \ra &=& - \frac{i M_N}{2} \, \la x \ra_{\delta u-\delta d} \,,
\label{eq:me2momentT}
\ee
where $M_N$ is the nucleon mass.
The nucleon interpolating operator ${\mathcal N}$ used is 
\be
{\mathcal N} = \epsilon^{abc} \Big[  q_1^{aT} (x) C \gf \frac{(1\pm \gamma_4)}{2}q_2^b(x) \Big] q^c_1(x) \,,
\label{nucop}
\ee
where $\{a,b,c\}$ are color indices,
$q_1,q_2 \in \{u,d\}$ and $C=\gamma_0 \gamma_2$ is the charge conjugation matrix.  The nonrelativistic projection $(1\pm \gamma_4)/2 $ is inserted to improve the signal, with the plus and minus signs applied to the forward and backward propagation in Euclidean time,
respectively. At zero momentum, this operator couples only to the spin $\frac{1}{2}$ states. The zero momentum two-point and three-point nucleon correlation functions are defined as
\begin{flalign}
\bold{C}^{2pt}_{\alpha \beta} (\tau ) &= \sum_{\bm x}\la 0 | {\mathcal N}_\alpha (\tau, {\bm x}) \ol{{\mathcal N}}_\beta(0,{\bm 0})| 0\ra\\
\bold{C}^{3pt}_{\mathcal{O},\alpha \beta}
(\tau,t ) &= \sum_{{\bm x}',{\bm x}} \la 0 | {\mathcal N}_\alpha (\tau, {\bm x}) {\cal O} (t, {\bm x}') \ol{{\mathcal N}}_\beta(0,{\bm 0})| 0\ra
\end{flalign}
where $\alpha$, $\beta$ are spin
indices. The source is placed at time slice 0, the sink is at $\tau$
and the one-derivative operators
inserted at time slice $t$.  Data have been accumulated for the
values of $\tau$ specified in Tables~\ref{Table:HISQ-ensembles} and \ref{Table:Clover-ensembles}, and, in each
case, for all intermediate times $0 \leq t \leq \tau$.

\section{Controlling the excited state contamination}
A major challenge to precision results is removing the contribution of excited states in the three-point functions. These occur because the lattice nucleon interpolating operator,
couples to the nucleon, all its excitations and to multi particle states with the same quantum numbers. The strategy to remove these artifacts are described in Refs. \cite{Mondal:2020cmt, Mondal:2020ela}:
reduce ESC by using smeared sources in the generation of quark propagators and
then fit the data at multiple source-sink separations $\tau$ using the spectral decomposition of the correlation
functions keeping as many excited
states as possible without over-parameterizing
the fits.
The spectral decomposition of the zero-momentum two-point function,
${ C_{\rm 2pt}}$, truncated at four states, is given by
\begin{equation}
C_{\rm 2pt}(\tau) = \sum_{i=0}^3  |{\cal A}_i|^2e^{-M_i \tau} \,.
\label{eq:2pt}
\end{equation}
We fit the data over the largest time range,
$\{\tau_{min}$--$\tau_{max}\}$, allowed by statistics, i.e., by the
stability of the covariance matrix, to extract 
the masses $M_i$ and the amplitudes ${\cal A}_i$ for the creation/annihilation of the
four states by the interpolating operator.  We perform two types of four-state fits.  In the fit denoted $\{4\}$, we use the
empirical Bayesian technique described in the Ref.~\cite{Yoon:2016jzj}
to stabilize the three excited-state parameters. In the second fit,
denoted $\{4^{N\pi}\}$, we use a normally distributed prior for $M_1$,
centered at the lower of the non-interacting energy of $N({-\bm
1}) \pi({\bm 1})$ or the $N({\bm 0})\pi({\bm 0}) \pi({\bm 0})$
state, and with  a width of 0.04--0.05 in lattice
units. 

In the fits to the two-point functions, the $\{4\}$ and $\{4^{N\pi}\}$ strategies cannot be distinguished on the basis of the $\chi^2/$dof.
In fact, the full range of $M_1$ values between the two estimates, from $\{4\}$ and $\{4^{N\pi}\}$, are viable on the basis of
$\chi^2/$dof alone. The same is true of the values for $M_2$, indicating a large flat region in parameter space.  Because of this
large region of possible values for the excited-state masses, $M_i$, we carry out the full analysis with three strategies that use
different estimates of $M_1$ and investigate the sensitivity of the results on them.

The analysis of the three-point functions,
$C_\mathcal{O}^{3\text{pt}}$, is performed retaining up to three states $|i\rangle$ in the spectral decomposition:
\begin{equation}
  C_\mathcal{O}^{3\text{pt}}(\tau;t) =
   \sum_{i,j=0}^2 \abs{\mathcal{A}_i} \abs{\mathcal{A}_j}\matrixe{i}{\mathcal{O}}{j} e^{-M_i t - M_j(\tau-t)}\,.
   \label{eq:3pt}
\end{equation}
To remove the ESC and extract the desired ground-state matrix element, $\matrixe{0}{\mathcal{O}}{0}$, we make a simultaneous fit in $t$ and $\tau$.  In choosing the set of points,
$\{t,\tau\}$, to include in the final fit, we attempt to balance statistical and systematic errors. First, we neglect $t_{\rm skip}$
points next to the source and sink in the fits as these have the largest ESC. Next, noting that the data at smaller $\tau$ have exponentially smaller
errors but larger ESC, we pick the largest three values of $\tau$ for all seven ensembles.
Since errors in the data grow with
$\tau$, we partially compensate for the larger weight given to smaller $\tau$ data by choosing $t_{\rm skip}$ to be the same for all $\tau$,
i.e., by including increasingly more $t$ points with larger $\tau$, the weight of the larger $\tau$ data points is increased.  Most of our
analysis uses a $3^\ast$-fit, which is a three-state fit with the term containing $\langle 2 | {\mathcal O} | 2\rangle$ set to zero, as it is undetermined and its inclusion results in an overparameterization based on the Akaike information criteria~\cite{akaike}.

\begin{figure}[htbp]
\begin{subfigure}
\centering
\includegraphics[angle=0,width=0.32\textwidth]{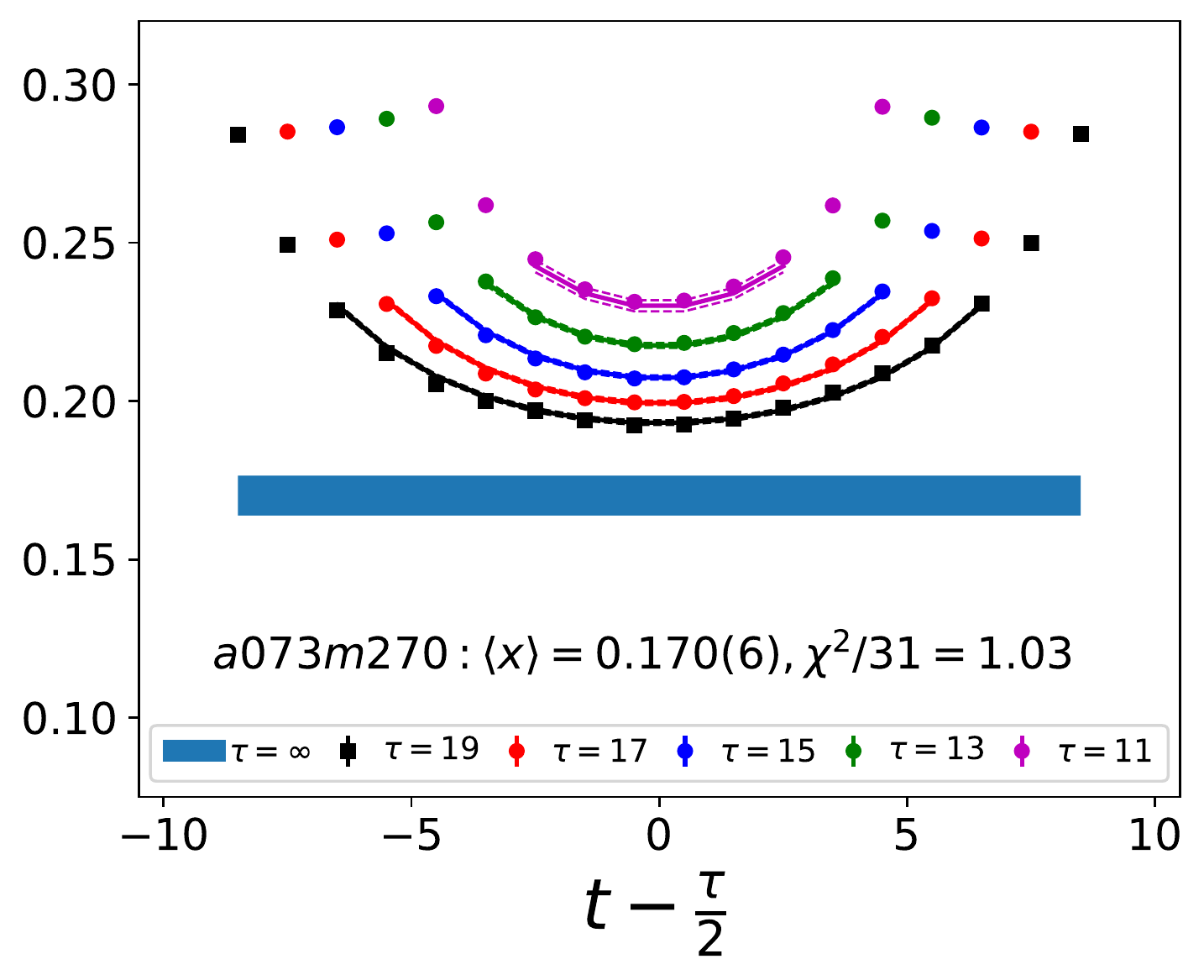}
\includegraphics[angle=0,width=0.32\textwidth]{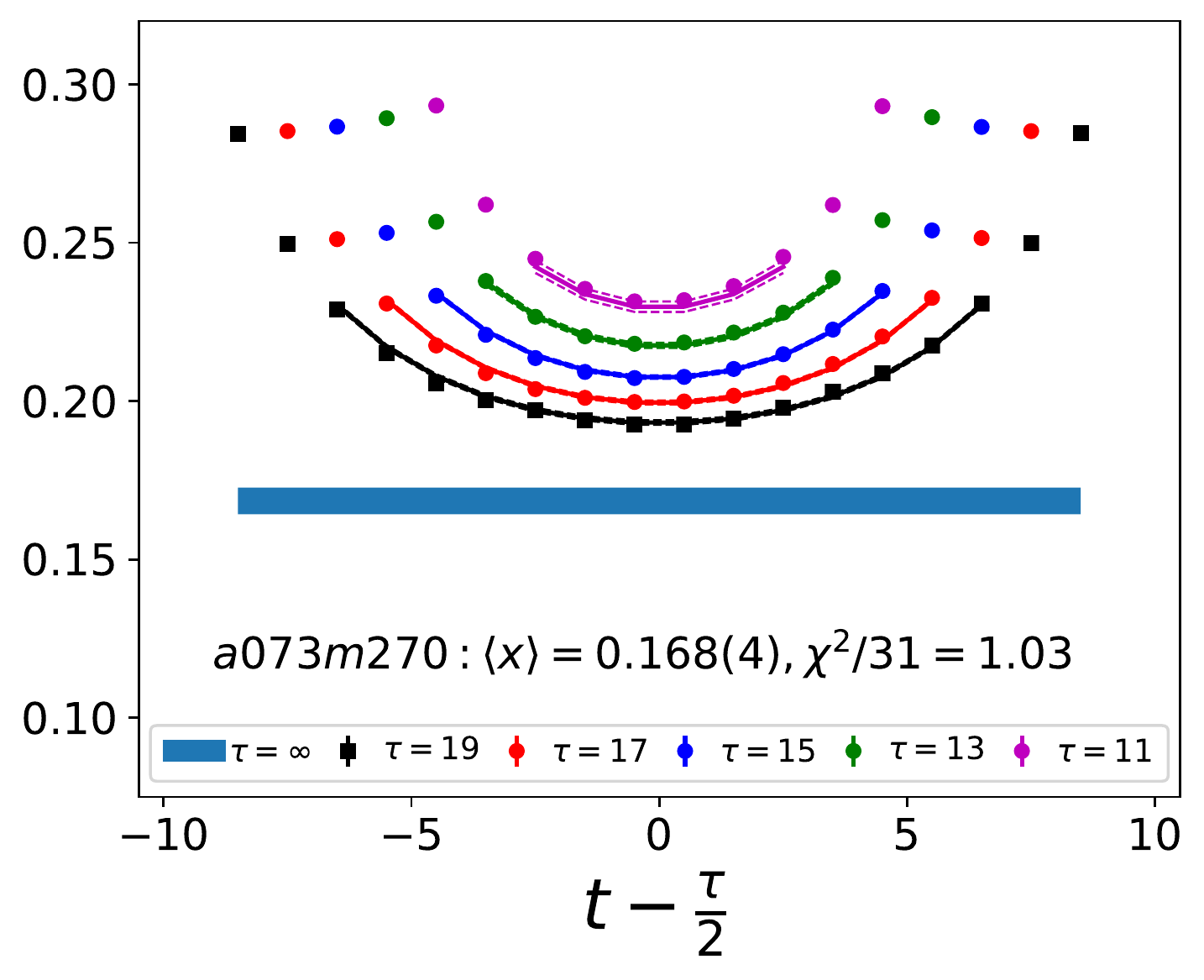}
\includegraphics[angle=0,width=0.32\textwidth]{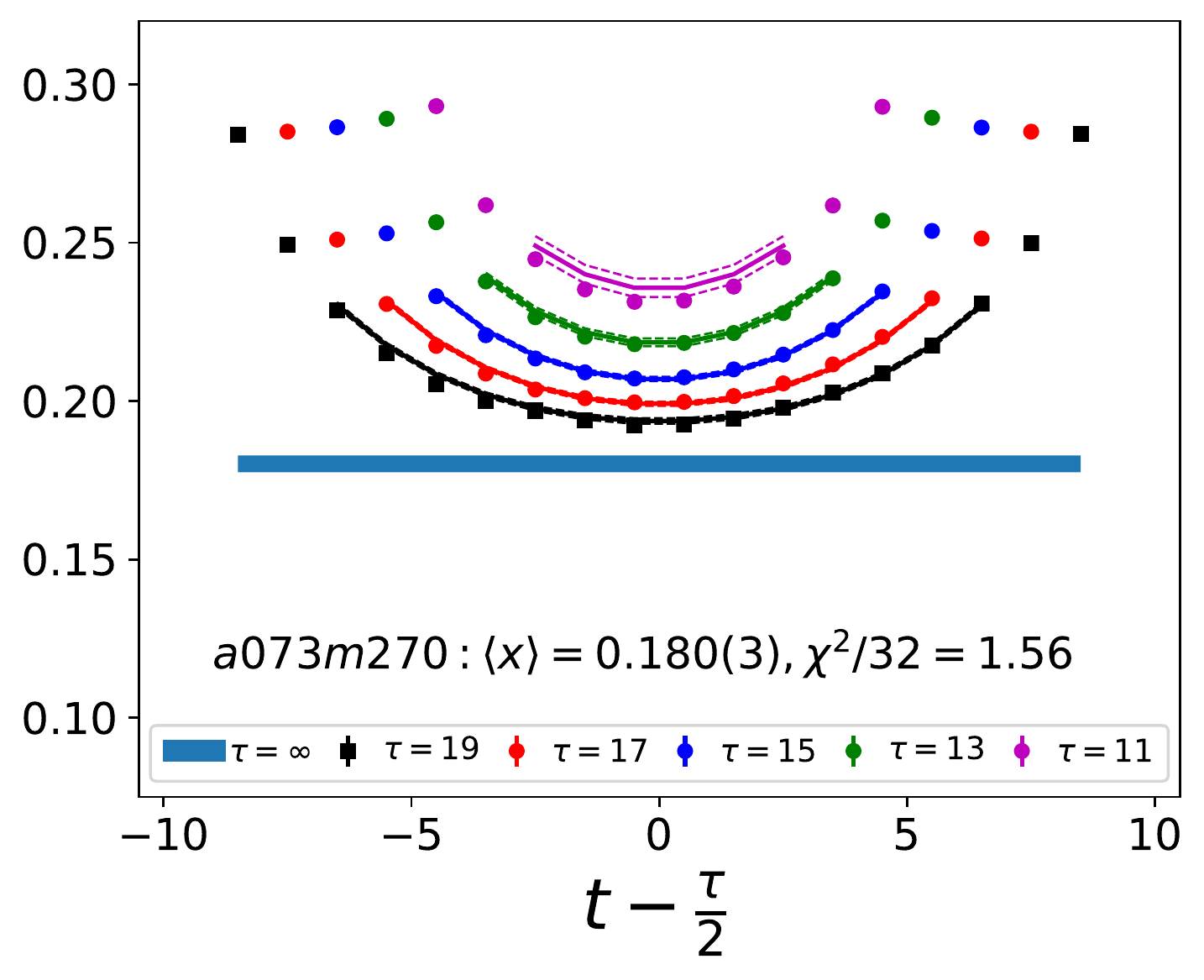}
\end{subfigure}
\vspace{-0.3cm}
\caption{Data of the ratio $\frac{C_{\cal O}^{3pt}(\tau,t )}{C^{2pt}(\tau)}$ scaled with the kinematic factors to give $\la x \ra_{u-d}$  for the ensemble $a073m270$.
The three panels show fits to the data with the largest three values of $\tau$ using three strategies:
$\{ 4, 3^*\}$  (left panel), $\{ 4^{N\pi}, 3^*\}$  (middle panel) and $\{ 4, 2^{\rm free}\}$ (right panel). The fits are performed using data for the largest three values of $\tau$. For each $\tau$ , the line in the same color as the data points is the result of the fit used to obtain the ground state matrix element.
The blue band in each plot indicate the value of the moment obtained via the ground state matrix element from the fit.  
}
\label{Fig:ratio-plots}
\end{figure}

To investigate the sensitivity of
$\matrixe{0}{\mathcal{O}}{0}$ to possible values of $M_i$ we carry out the full analysis with three strategies using the mnemonic $\{m,n\}$ to denote an $m$-state fit to the
two-point function and an $n$-state fit to the three-point function. Figure~\ref{Fig:ratio-plots} shows an example of difference in estimates from the three fit strategies 
for $\la x \ra_{u-d}$ from  $a073m270$.
\begin{itemize}
\item
$\{4,3^*\}$: The spectrum is taken from a $\{4\}$ state
fit to the two-point function using Eq.~\eqref{eq:2pt} and then a
$\{3^\ast\}$ fit is made to the three-point function using
Eq.~\eqref{eq:3pt}. Both fits are made within a single jackknife loop.
This is the standard strategy, which assumes that the same set of
states are dominant in the two- and three-point functions.
\item
$\{4^{N\pi},3^*\}$: The excited state spectrum is taken from a four-state
fit to the two-point function but with a narrow prior for the first
excited state mass taken to be the energy of a non-interacting $N
({\bm p}=1) \pi({\bm p}=-1)$ state (or $N ({\bm0}) \pi({\bm0}) \pi({\bm0})$ that has
roughly the same energy).  This spectrum is then used in a
$\{3^\ast\}$ fit to the three-point function. This variant of the
$\{4,3^*\}$ strategy assumes that the lowest of the theoretically
allowed tower of $N \pi$ (or $N \pi \pi $) states contributes.
\item
$\{4,2^{\rm free}\}$: The only parameters taken from the $\{4\}$ state
fit are the ground state amplitude ${\cal A}_0$ and mass $M_0$, whose
determination is robust. In the
two-state fit to the three-point function, the mass of the first excited state, $M_1$, is left as a free parameter, ie, the most
important determinant of ESC, $M_1$, is obtained from the fit to the three-point function.  The relative limitation of the $\{4,2^{\rm free}\}$
strategy is that, with the current data, we can only make two-state fits to the three-point functions, i.e., include only one excited state.
\end{itemize}

\section{Chiral, continuum and finite volume (CCFV) extrapolations}
\begin{figure}
\centering
\begin{subfigure}
\centering
\includegraphics[angle=0,width=0.32\textwidth]{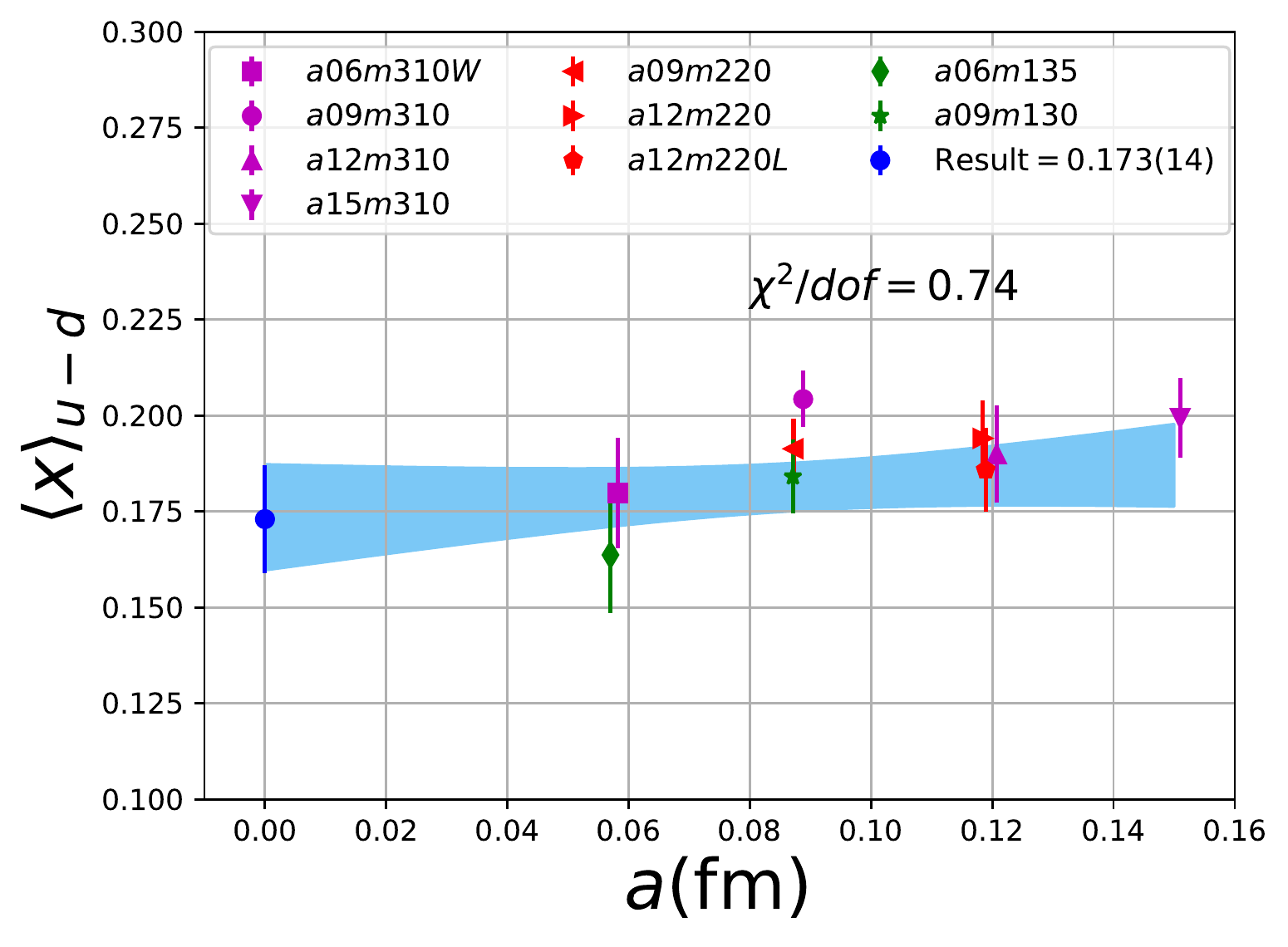}
\end{subfigure}
\begin{subfigure}
\centering
\includegraphics[angle=0,width=0.32\textwidth]{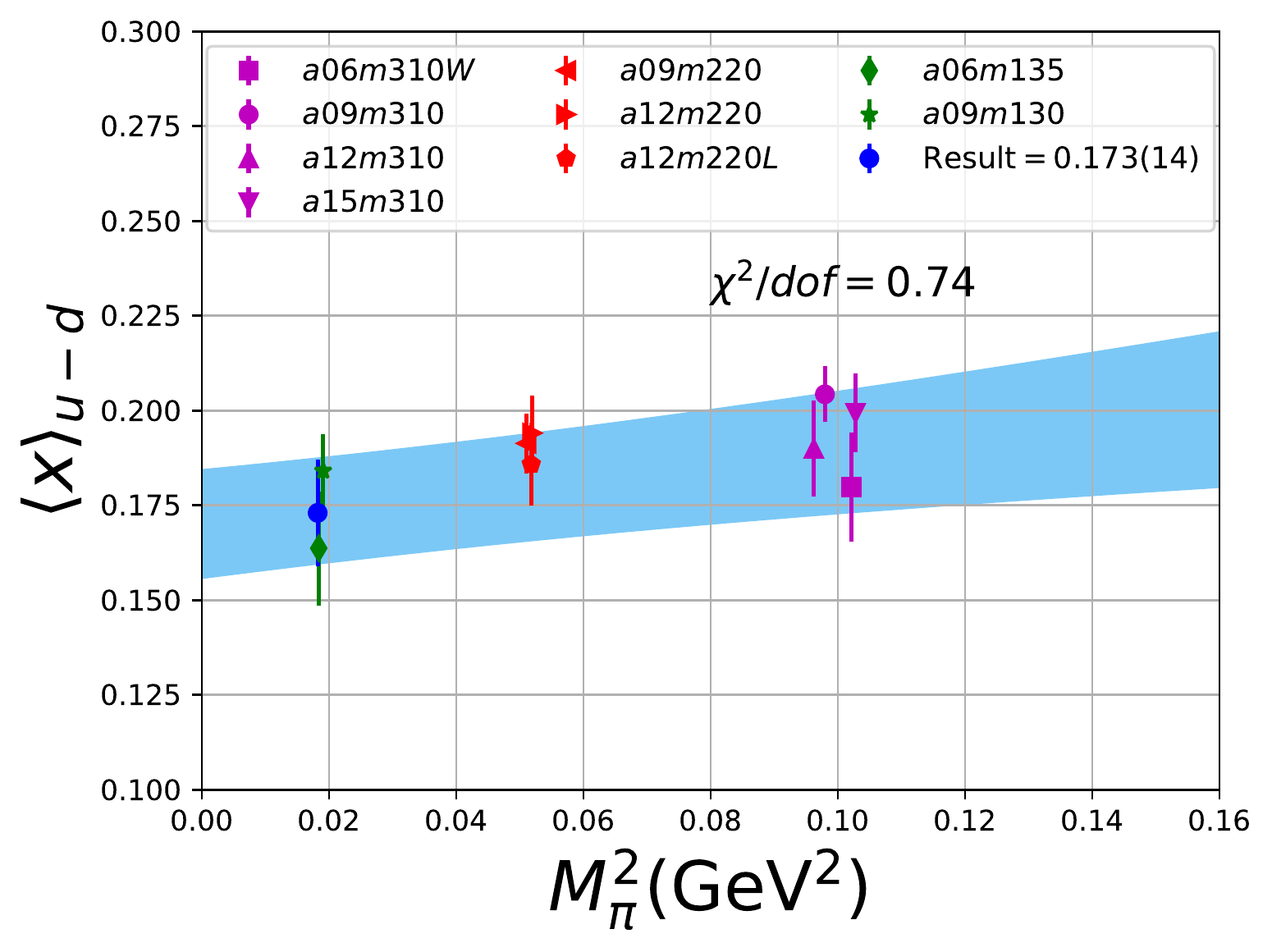}
\end{subfigure}
\caption{Clover-on-HISQ data for $\la x \ra_{u-d}$ obtained via the fit strategy $\{ 4, 3^* \}$, renormalized in the ${\ol {MS}}$ scheme at $\mu = 2$ GeV, for all nine ensembles (PNDME 20). The blue band  shows the CC fit result. In the left panel the fit is evaluated at $M_\pi = 135$ MeV and plotted versus $a$, while in the right panel it shows the result versus $M_\pi^2$
evaluated at $a = 0$.}
 \label{Fig:CC-PNDME}
\end{figure}

To obtain the final, physical results at $M_\pi=135$~MeV, $M_\pi L \to
\infty$ and $a=0$, we make a simultaneous CCFV fit of data renormalized in the ${\ol MS}$ scheme at $\mu = 2$ GeV keeping only the
leading correction term in each variable:
\begin{flalign}
\la x \ra (M_\pi; a;L) &= c_1 + c_2 a +c_3 M_\pi^2
+ c_4 \frac{M_\pi^2~ e^{-M_\pi L}}{\sqrt{M_\pi L}} \,.
\label{eq:CCFV}
\end{flalign}
Note that, in both lattice setups the discretization errors start with a term linear in $a$. The results of the CCFV fits in case of PNDME 20 show that the finite volume correction term, $c_4$, is not constrained. Therefore, for PNDME 20, we use CC fits (i.e., with $c_4=0$ in Eq. \ref{eq:CCFV}) to obtain the final results.
As an example, in Fig. \ref{Fig:CC-PNDME} we present the  PNDME 20 CC fit to $\la x \ra_{u-d}$ obtained via $\{ 4, 3^* \}$ fit strategy.

\begin{figure}
\centering
\begin{subfigure}
\centering
\includegraphics[angle=0,width=0.32\textwidth]{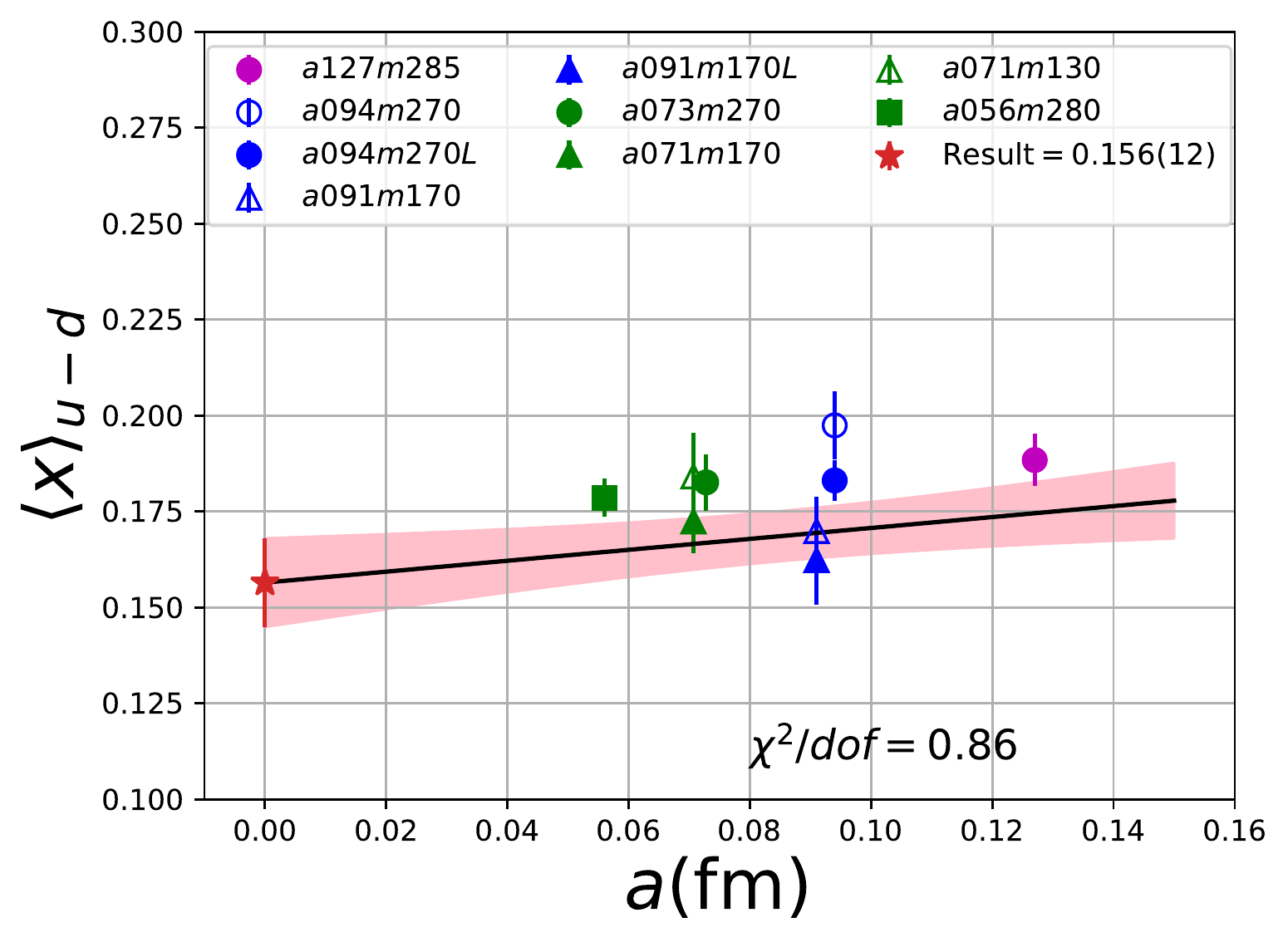}
\end{subfigure}
\begin{subfigure}
\centering
\includegraphics[angle=0,width=0.32\textwidth]{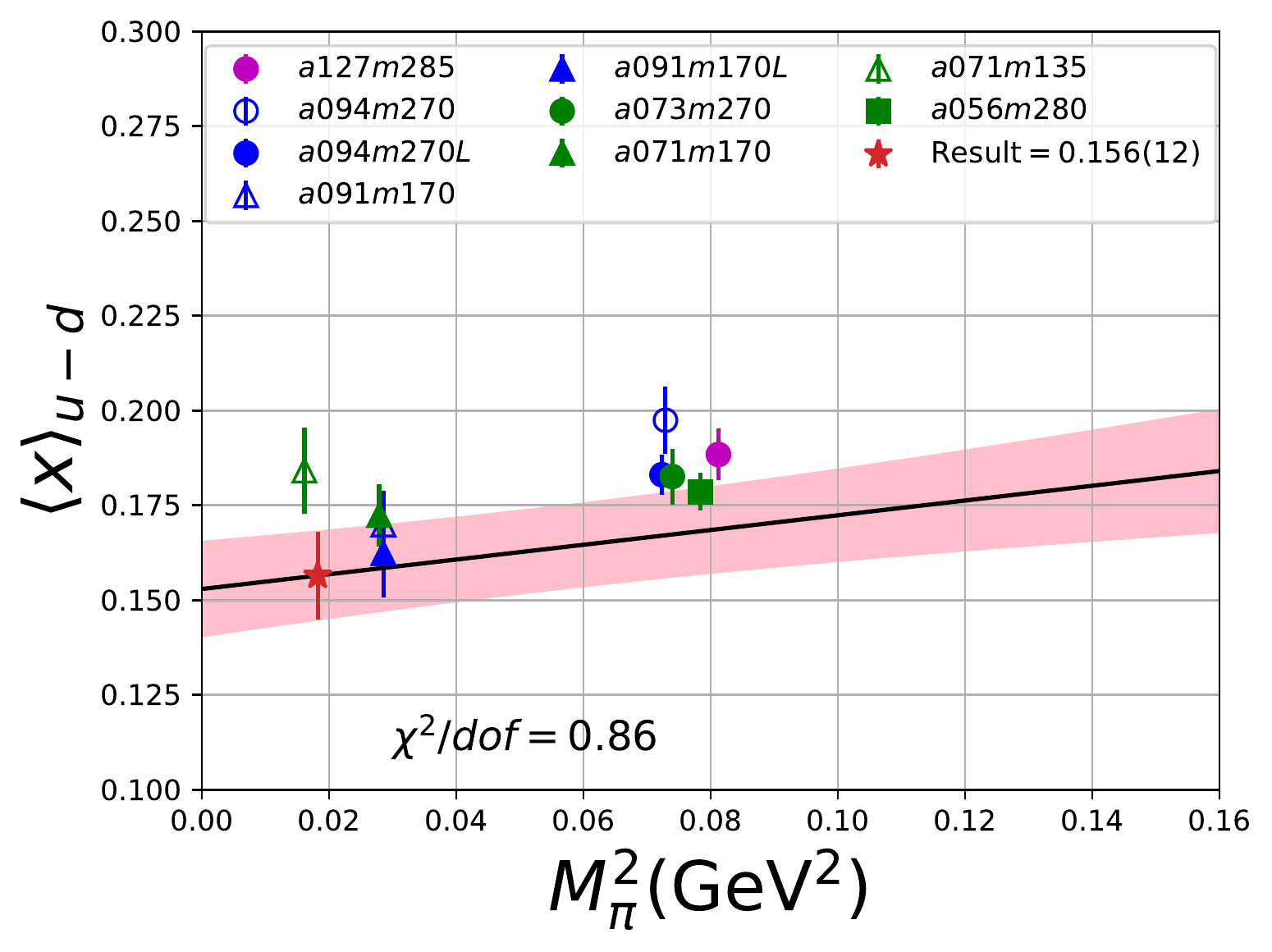}
\end{subfigure}
\begin{subfigure}
\centering
\includegraphics[angle=0,width=0.32\textwidth]{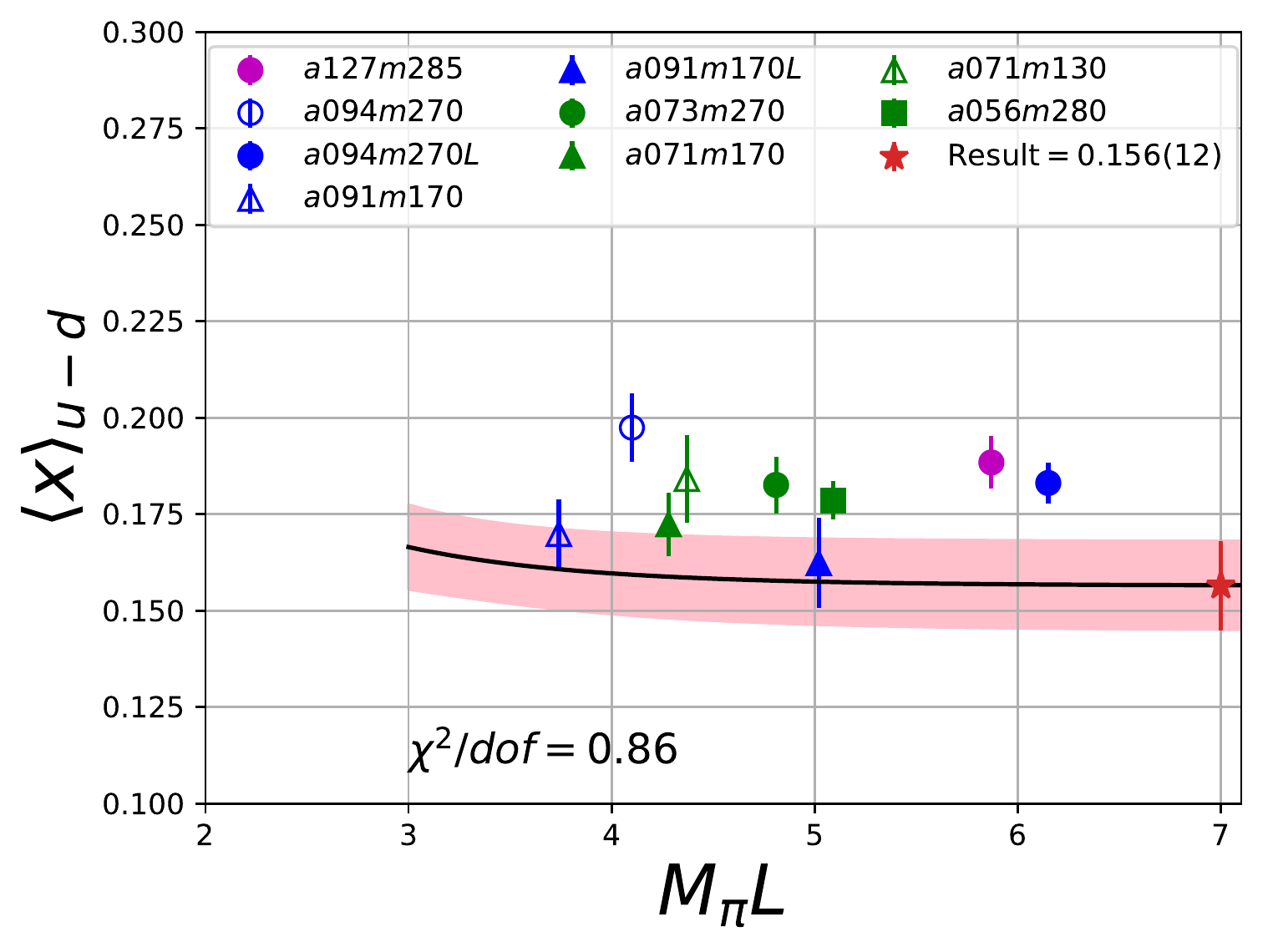}
\end{subfigure}
\caption{Clover-on-Clover data for $\la x \ra_{u-d}$ obtained via the fit strategy $\{ 4, 3^* \}$, renormalized in the ${\ol MS}$ scheme at $\mu = 2$ GeV, for seven ensembles ensembles (first row, NME 20) and for nine ensembles (second row, NME 21). The pink band shows the CCFV fit result. (Left) result evaluated at $M_\pi = 135$ MeV and $M_\pi L = \infty$ and plotted versus $a$. (Middle) result plotted versus $M_\pi^2$ and evaluated at $a = 0$ and $M_\pi L = \infty$. (Right) result plotted versus $M_\pi L$ and evaluated at $a = 0$ and $M_\pi  = 135$ MeV }
\label{Fig:CCFV-NME}
\end{figure}

The NME 20 and 21 data are sensitive to the finite volume corrections. Therefore, we use CCFV fits to obtain the final results as shown in Fig.~\ref{Fig:CCFV-NME} for $\la x \ra_{u-d}$ obtained via $\{ 4, 3^* \}$ fit strategy.

In both lattice formulations and for all three moments, we find only a small positive slope with respect to both the lattice spacing and $M_\pi$. The main difference between the PNDME 20 and the NME 20/NME 21 results is the $\sim 10\%$ decrease due to the finite volume correction in the latter. 

\begin{figure}
\centering
\begin{subfigure}
\centering
\includegraphics[angle=0,width=0.32\textwidth]{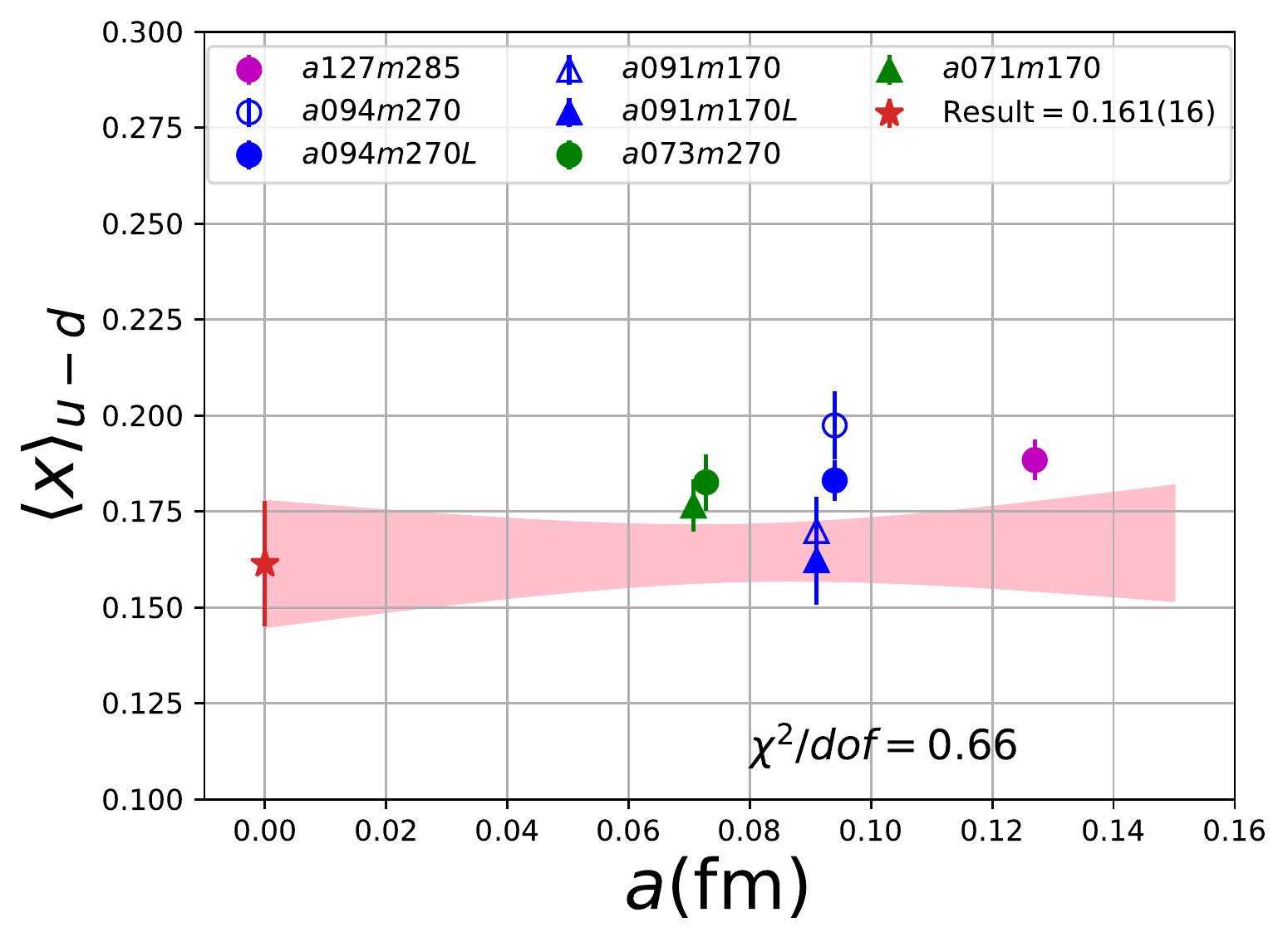}
\end{subfigure}
\begin{subfigure}
\centering
\includegraphics[angle=0,width=0.32\textwidth]{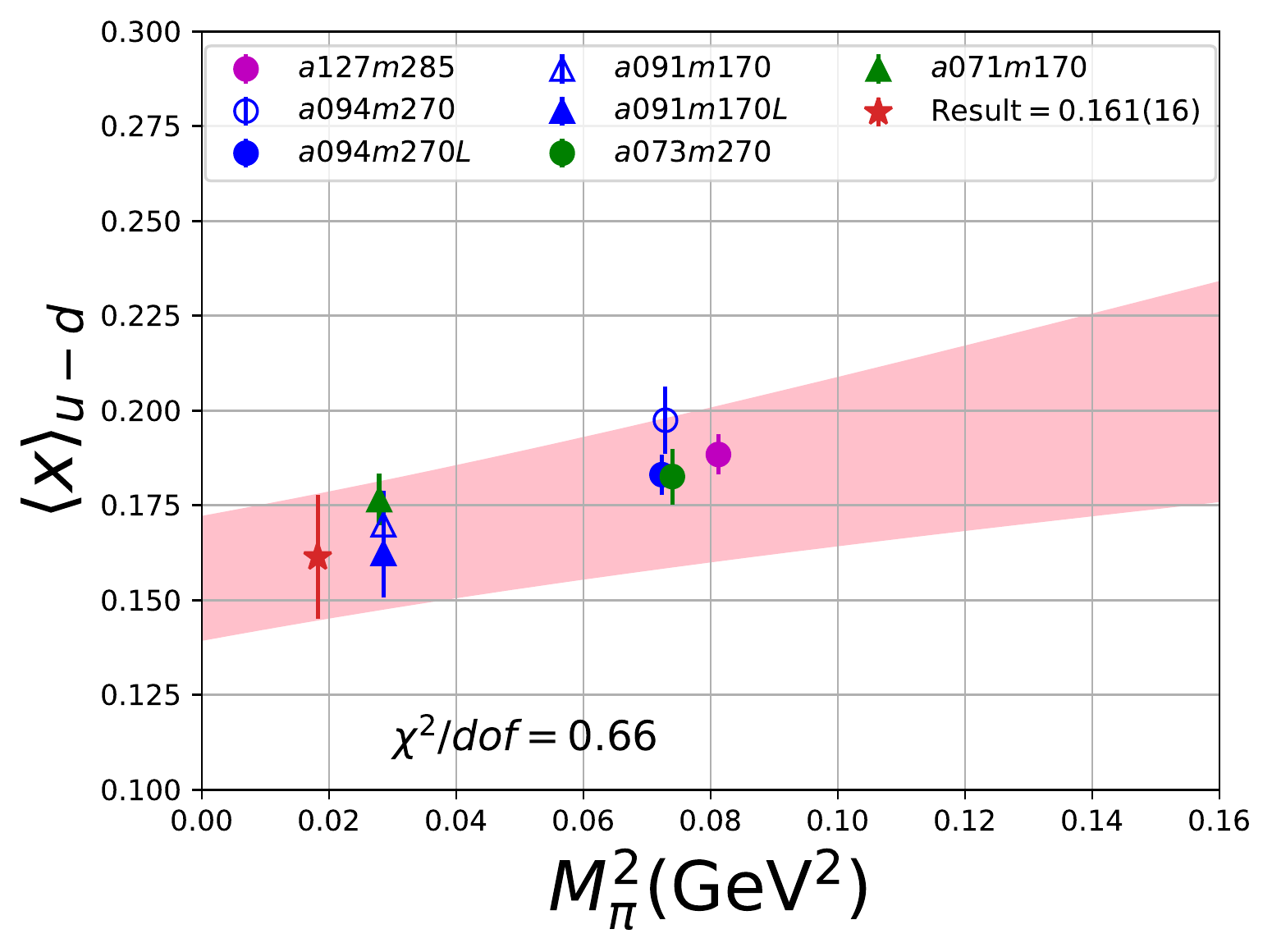}
\end{subfigure}
\begin{subfigure}
\centering
\includegraphics[angle=0,width=0.32\textwidth]{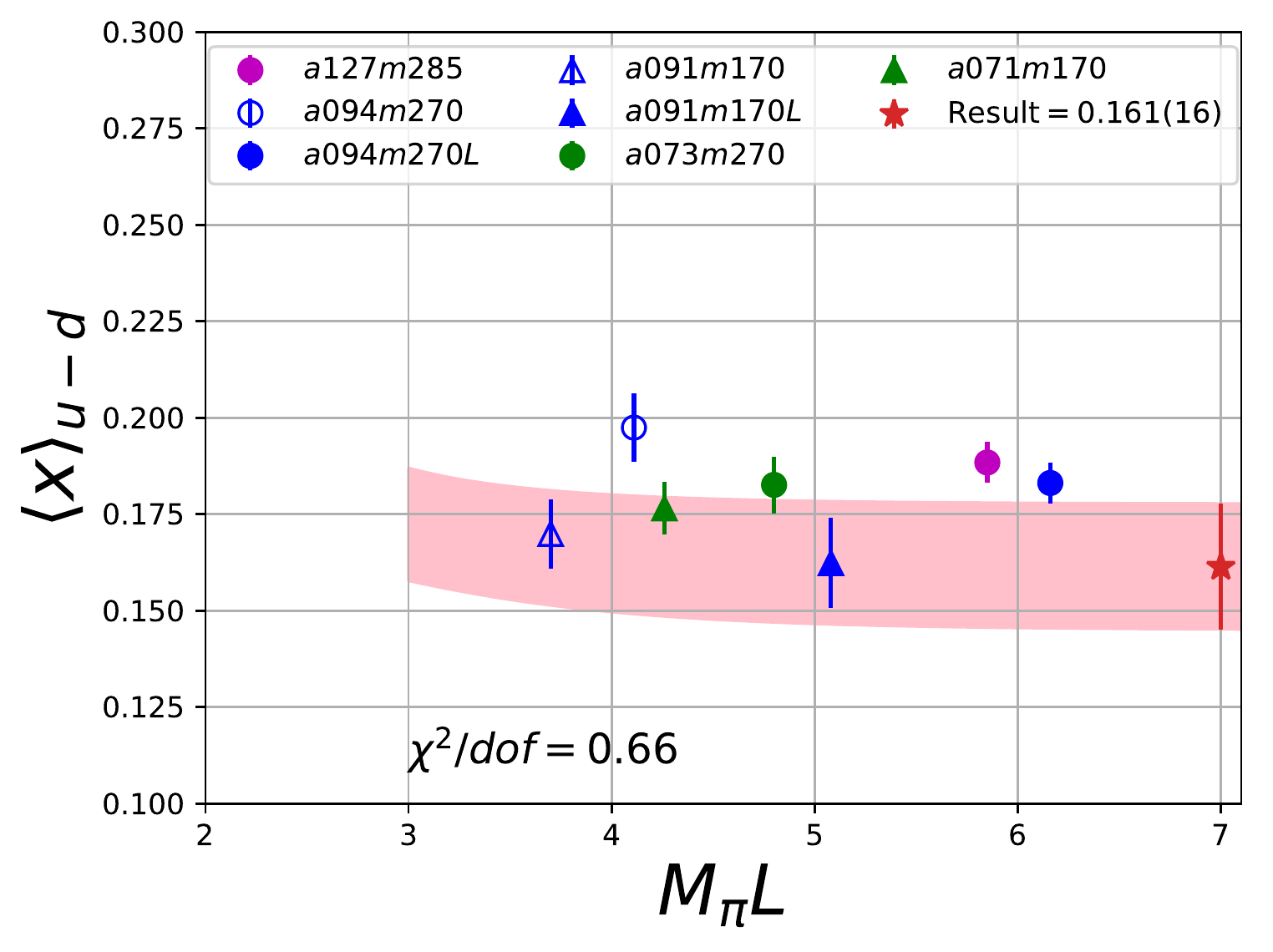}
\end{subfigure}
\end{figure}

\section{Results, comparison with the world data and conclusions}
\begin{table}[htbp]
\centering
\setlength{\tabcolsep}{0.6pt}
\renewcommand{\arraystretch}{1.1}
\begin{tabular}{|c|c|c|c| }
\hline

Moment & PNDME 20&NME 20 & NME 21\\
\hline
\hline
$\langle x \rangle_{u -d}$               &~ 0.173(14)(07)~ & ~0.155(17)(20)~   &
~0.156(12)(20)~          \\
\hline\hline
$\langle x \rangle_{\Delta u- \Delta d}$ & 0.213(15)(22) & 0.183(14)(20)   & 0.185(12)(20)          \\
\hline\hline
$\langle x \rangle_{\delta u- \delta d}$ & 0.208(19)(24) & 0.220(18)(20)   & 0.209(15)(20)          \\
\hline\hline
\end{tabular}
\caption{Final results for the moments.}
\label{Table:comparison-pndme-nme}
\end{table}
In Table \ref{Table:comparison-pndme-nme} we compare PNDME 20, NME 20 and NME 21 results. We choose the results from CCFV/CC fits of the moments obtained via $\{ 4, 3^*\}$ fit strategy of the correlators. The first number inside brackets is overall statistical error. We will also take
half the spread in results between $\{ 4^{N\pi}, 3^*\}$ and the $\{4,2^{\rm free}\}$ fit strategies as a second uncertainty (the second number inside the brackets) to account for possible unresolved
bias from incomplete control over  ESC. 

Our NME results are consistent with the PNDME 20 results. This is a valuable check of the PNDME 20 calculation that uses the non-unitary clover-on-HISQ lattice formulation.
For $\la x \ra_{u-d}$ and $\la x \ra_{\Delta u-\Delta d}$ the NME results are $\approx 1\sigma$ smaller than the PNDME. A large part of the difference is due to the finite-volume correction in the NME results.
There is reduction of statistical errors for all three quantities on going from NME 20 to NME 21.  This is due to adding data from two new ensembles $a071m130$ and $a056m280$ in NME 21 which gives larger ranges in both the lattice-spacing and the pion mass in the CCFV fits.

\begin{figure}
\centering
\begin{subfigure}
\centering
\includegraphics[angle=0,width=0.29\textwidth]{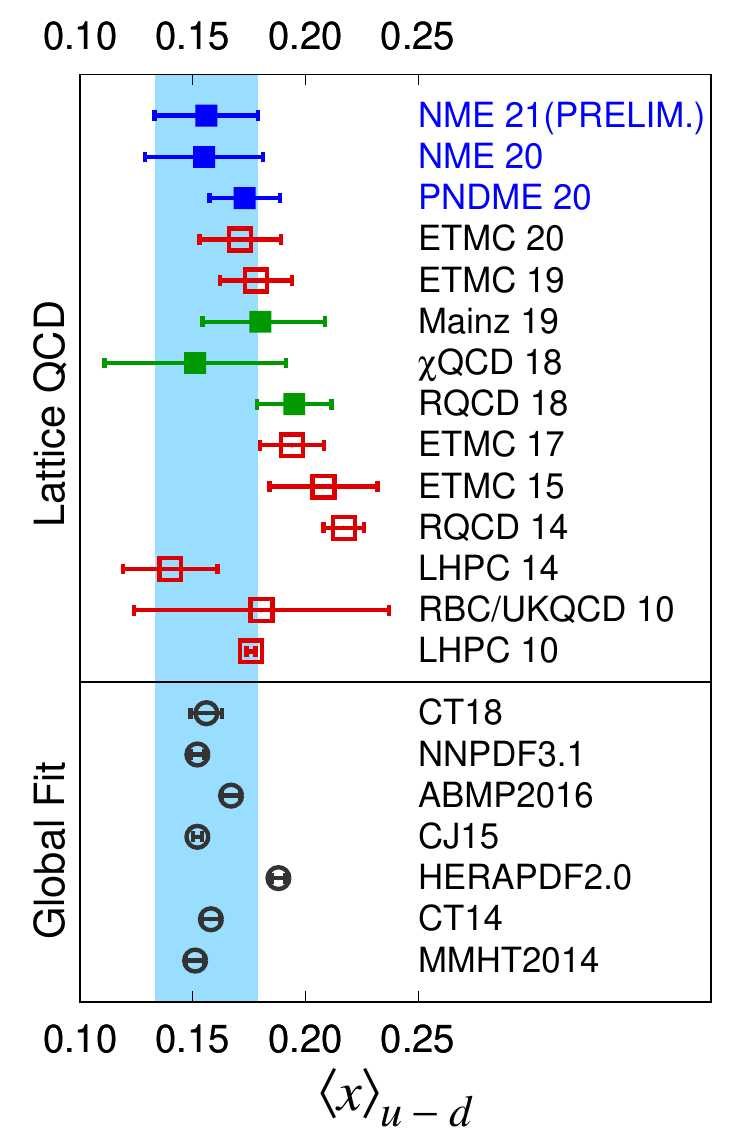}
\end{subfigure}
\begin{subfigure}
\centering
\includegraphics[angle=0,width=0.29\textwidth]{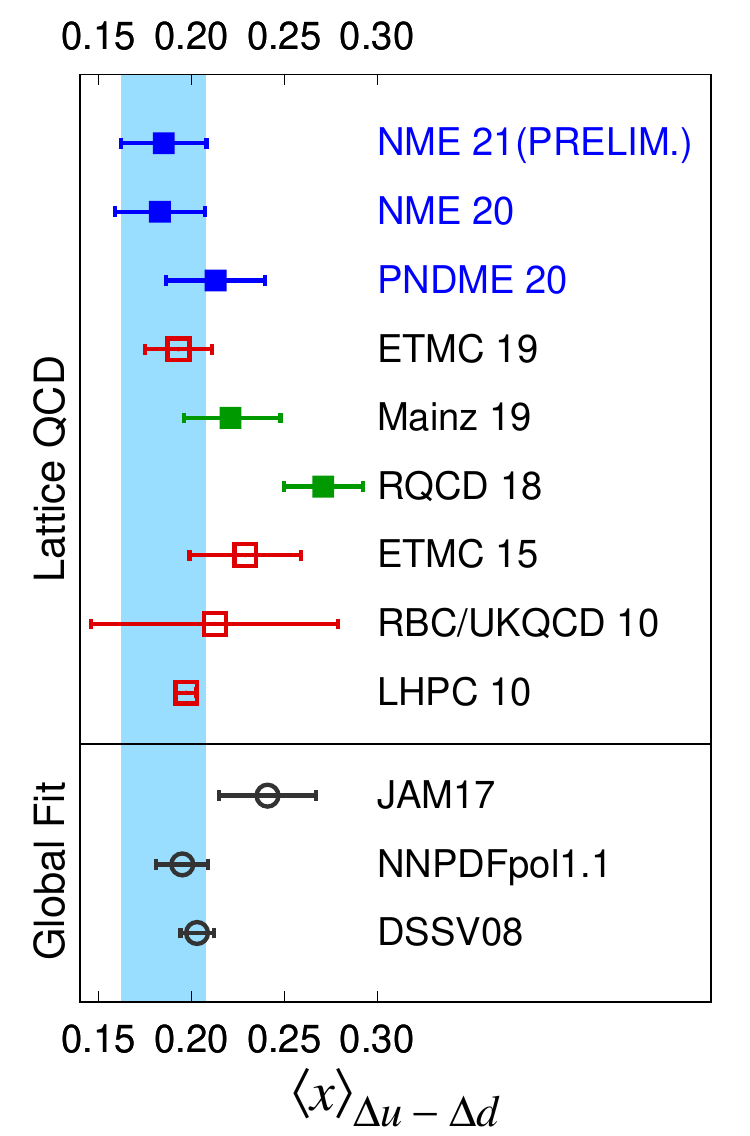}
\end{subfigure}
\begin{subfigure}
\centering
\includegraphics[angle=0,width=0.29\textwidth]{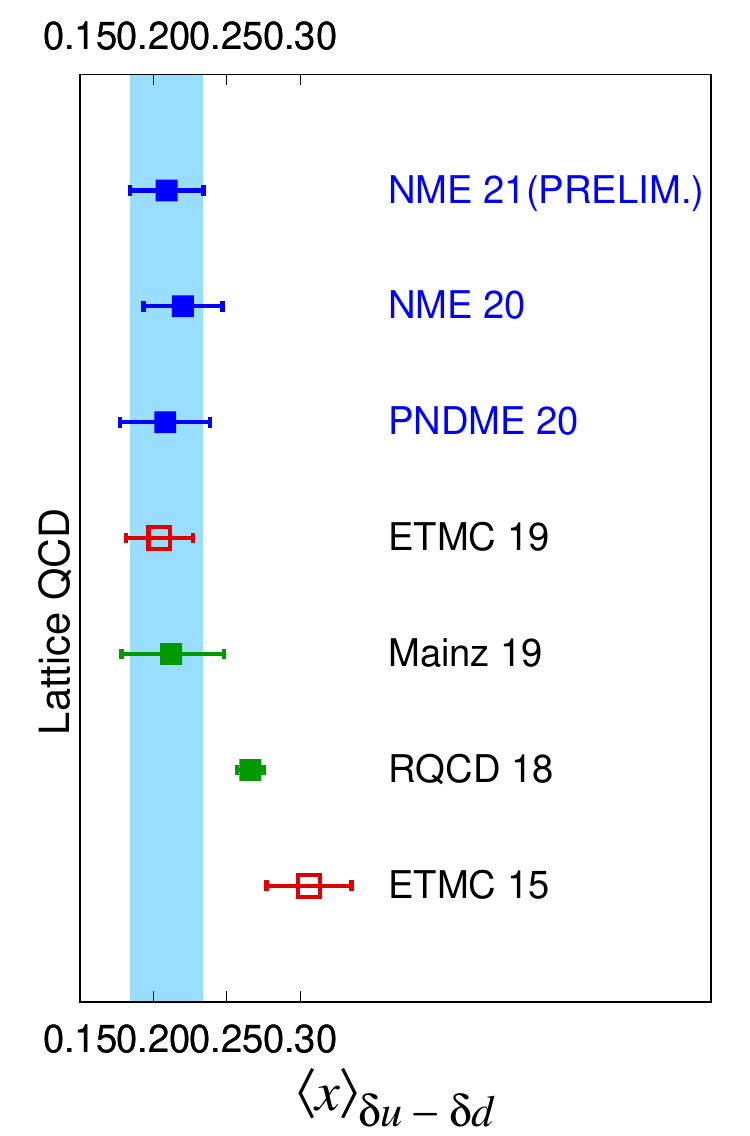}
\end{subfigure}
\caption{A comparison of results from lattice QCD calculations with dynamical fermions and global fits (below the black line). The left panel compares results for the momentum fraction, the middle for the helicity moment, and the right for the transversity moment. Our NME 21 result (preliminary) is also shown as the blue band to facilitate comparison.}
\label{Fig:comparison-world-data}
\end{figure}

An updated comparison of our results with other lattice calculations
and phenomenological global fit estimates is given in Fig. \ref{Fig:comparison-world-data}. They are in good agreement
with other recent lattice results by ETMC \cite{Alexandrou:2020sml, Alexandrou:2019ali}, Mainz \cite{Harris:2019bih} and $\chi$QCD \cite{Yang:2018nqn} collaborations.
Our estimate for the momentum fraction is in good agreement with most global fit estimates but has much larger error. The three estimates for the
helicity moment from global fits have a large spread, and our estimate is consistent with the smaller error
estimates. Lattice estimates for the transversity moment are a prediction.

\section{Acknowledgements}
We thank the MILC Collaboration for sharing the
HISQ ensembles. The calculations used the Chroma software
suite~\cite{Edwards:2004sx}. We gratefully acknowledge
computing resources provided by NERSC, OLCF at Oak Ridge, USQCD and
LANL Institutional Computing. Support for this work was provided the
U.S. DOE Office of Science, HEP and NP, the  NSF, and by the LANL LDRD program. The work of SM and HL are partially supported by the US National Science Foundation under grant PHY 1653405 ``CAREER: Constraining Parton Distribution Functions for New-Physics Searches'' and by the  Research  Corporation  for  Science  Advancement through the Cottrell Scholar Award.

\end{document}